\documentstyle[12pt,psfig,aaspp4]{article}
\newcommand\etal{{et al.\/}}
\begin{document}

\title{Determination of the Baryon Density from Large Scale Galaxy Redshift
Surveys}

\author{David M. Goldberg and Michael A. Strauss\altaffilmark{1}}
\authoremail{(goldberg,strauss)@astro.princeton.edu}
\affil{Princeton University Observatory, Princeton, NJ 08544-1001}

\altaffiltext{1}{Alfred P. Sloan Foundation Fellow}

\begin{abstract}
We estimate the degree to which the baryon density,  $\Omega_{b}$,  can be 
determined from the galaxy power spectrum measured from large
scale galaxy redshift surveys, and in particular, the Sloan Digital Sky 
Survey.  A high baryon density will cause wiggles to appear in the 
power spectrum, which should be observable at the current epoch.  We
assume linear theory on scales $\geq 20h^{-1}{\rm Mpc}$
and do not include the effects of redshift distortions, evolution, or biasing.
With an optimum estimate of $P(k)$ to $k\sim 2\pi/(20 h^{-1} {\rm Mpc})$, the
$1\,\sigma$ uncertainties in $\Omega_{b}$ are roughly $0.07$ 
and $0.016$ in flat
and open ($\Omega_{0}=0.3$) cosmological models, respectively.  This result
suggests that it should be possible to test for consistency with 
big bang nucleosynthesis estimates of $\Omega_{b}$ if we live in an
open universe. 
\end{abstract}


\section{Introduction}

In anticipation of the forthcoming Sloan Digital Sky Survey (hereafter SDSS;
see {\tt http://www.astro.princeton.edu/BBOOK}; Gunn \& Weinberg 1995; 
Strauss 1997), several 
papers have discussed the use of large redshift surveys to determine of the galaxy power
spectrum  (Vogeley \& Szalay 1996; Tegmark \etal\ 1997ab; Hamilton
1997ab) and parameters derived therefrom (Tegmark 1997; Nakamura,
Matsubara, \& Suto 1997; de Laix \& Starkman 1997). 
This paper will quantify how well we can constrain the baryon density,
$\Omega_{b}$, from the power spectrum of the SDSS redshift survey.

The universal ratio of baryonic mass density to critical mass density, 
$\Omega_{b}$, is of
acute cosmological interest.  In the standard big bang model, the ratio of
baryon density to photon density, $\eta$, uniquely determines the relative
abundances of deuterium, $^3$He, $^4$He and $^7$Li.  
Although the number density of 
photons can be readily measured from the 
temperature and spectrum of the Cosmic Microwave Background 
Radiation, the measurement of baryon density is much more difficult.  
However, by determining the abundance of deuterium, for example, one can
compute a unique value of $\eta$ (Peebles 1993).  
From $\eta$ and the measured
photon density one can determine $\Omega_{b}$, up to uncertainties in 
the Hubble constant, $H_{0}$.  However, since deuterium is 
converted into $^{3}$He in stars, it is necessary to look at very old objects
in order to get an accurate estimate of the primordial deuterium abundance. 
Tytler, Fan, \& Burles (1996) measured the deuterium abundances in high redshift 
QSO absorption lines 
and determined $\Omega_{b}=0.024\pm 0.002h^{-2}$, where 
$h$=$H_{0}$/(100 km s$^{-1}$ Mpc$^{-1}$).  
We shall adopt a value of $h=0.65$ (Kundi\'{c} \etal\
1997) throughout this
paper, yielding a value of $\Omega_{b}=0.057\pm 0.005$ from the Tytler
\etal\ (1996) observations.

Can the baryon density be independently confirmed?  
The baryonic mass density can have a profound effect on the shape of the 
expected mass power spectrum on large scales. 
The baryonic component of the density field is 
strongly coupled to the radiation field until $z\simeq 1100$.   
There are fluctuations in the radiation power spectrum 
due to sound waves crossing the horizon at decoupling
(Padmanabhan 1995; Hu \& Sugiyama 1996).  Immediately
after decoupling, the baryon power spectrum is the same as that of the photons,
containing fluctuations, or ``wiggles'' on large 
scales.  If 
the matter field is almost entirely composed of Cold Dark Matter (CDM), 
however, the baryons
quickly fall into the dark matter potential wells, and much of these wiggles
are erased.  However, the greater the proportion of baryons,
$\Omega_b/\Omega_0$, the greater
the influence of the wiggles in the final (linearly evolving) power spectrum.

The change in shape of the power spectrum through decoupling has been modeled 
analytically by several authors.  $T(k)$
is defined as the transfer function 
of the power spectrum such that $P(k)\propto k^{n_{prim}} T(k)^{2}$, where 
$n_{prim}$ 
is the spectral index of the primordial power spectrum.
Holtzman (1989) computed $T(k)$ for a wide
variety of models, including 
some with non-zero values of $\Omega_{b}$.
Sugiyama (1995; see also
White \etal\ 1996; Peacock \& Dodds 1994) include a 
baryon correction
term to the analytic fit to the transfer function:
\begin{equation}
T(k)=\frac{\ln(1+2.34q)}{2.34q}\times
\left[1+3.89q+(16.1q)^{2}+(5.46q)^{3}+(6.71q)^{4}\right]^{-1/4}.
\label{eq:transfer} 
\end{equation}
Here, $q\equiv k/h\Gamma$, $\Omega_{0}$ is the ratio of mass density to the 
critical density, and
\begin{equation}
\Gamma=\Omega_{0}h 
\ \exp\left[-\Omega_{b}-\left(\frac{h}{0.5}\right)^{1/2}
\frac{\Omega_{b}}{\Omega_{0}}\right] \ .
\label{eq:Gamma} 
\end{equation}
This functional form is smooth; it does not contain
the wiggles discussed above and thus this analytic formulation gives 
results which are entirely degenerate for various values of $\Omega_{b}$ and
$\Omega_{0}$.  If we take the
wiggles into account, the degeneracy can be broken.

We use a numerical approach to quantify the effect of wiggles in
the power spectrum.
We use a package called CMBfast, written
by Seljak \& Zaldarraiga (1996) to calculate the
transfer function of the mass power spectrum for a wide variety of 
cosmological models.  

\begin{figure}
\centerline{\psfig{figure=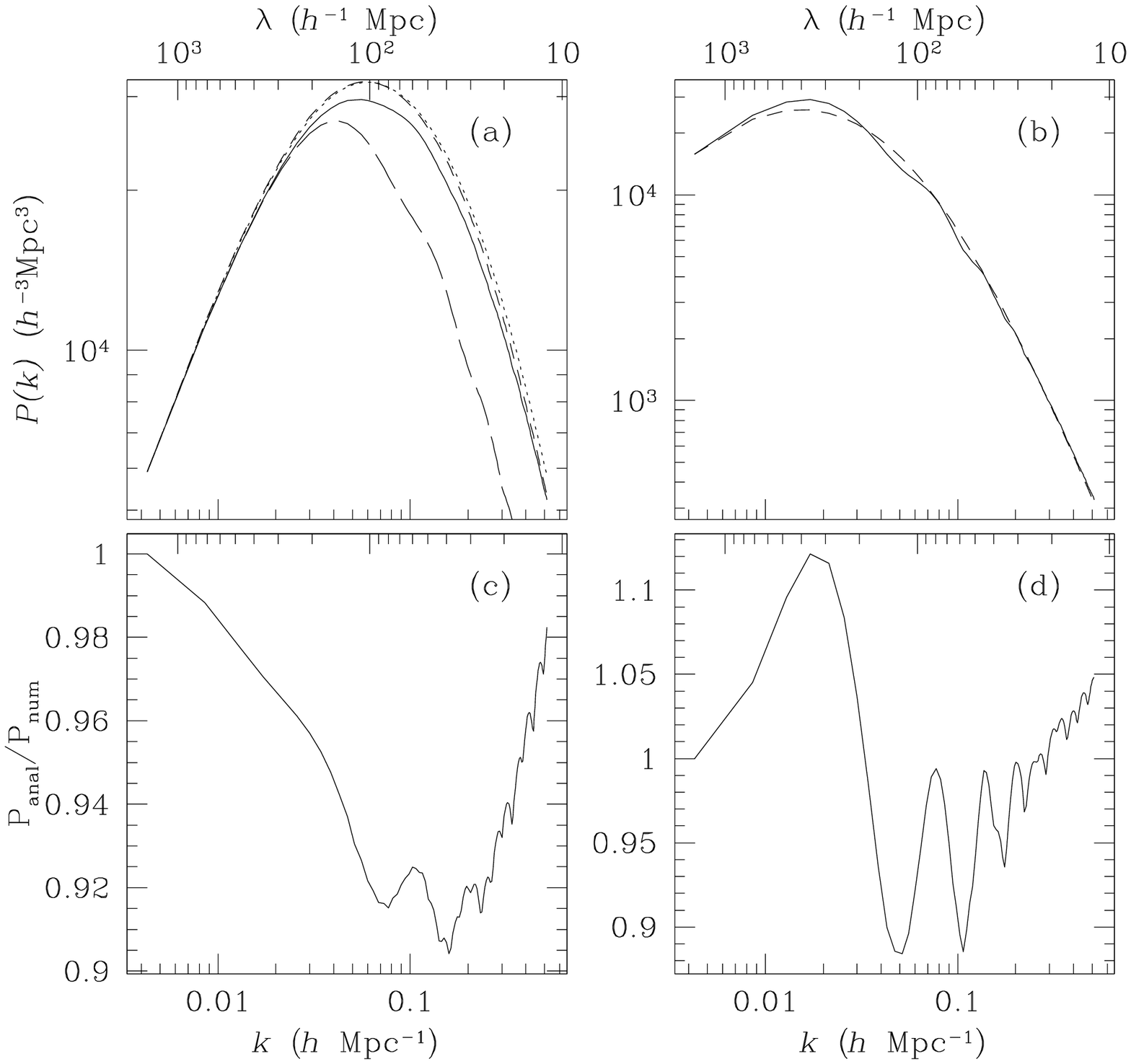,height=5.0in,angle=0}}
\caption{(a) Comparison of 4 power spectra, each with $h=0.65$, $n_{prim}=1$, 
and $\Gamma=0.58$. 
The solid line is the numerical result from CMBfast with 
$\Omega_{0}=1.0$,  $\Omega_{b}=0.057$.
The short-dashed line is the analytic power spectrum from 
equation~\ref{eq:transfer} using the same parameters. 
The dotted line is the numerical result 
for $\Omega_{0}=0.89$, $\Omega_{b}=0$.
The long-dashed line is the numerical 
result for $\Omega_{0}=1.37$, $\Omega_{b}=0.26$, corresponding to the same
ratio of $\Omega_{b}/\Omega_{0}$ as in the open model in (b).
(b) A comparison of the numerical (solid) and analytical (dashed) model power
spectra for an open cosmological model, with $\Omega_{0}=0.3$, 
$\Omega_{b}=0.057$, $\Gamma=0.173$. 
(c) The ratio of the analytic to numerical power spectra for
$\Omega_{0}=1$ and $\Omega_{b}=0.057$.
(d) The ratio of the analytic to numerical power spectra for $\Omega_{0}=0.3$
and $\Omega_{b}=0.057$.}
\label{fg:power}
\end{figure}

Figure~\ref{fg:power} illustrates the differences between the  
analytic fit of equation~(\ref{eq:transfer}) 
and the numerical approach.  In Figure~\ref{fg:power}a, 
we plot four power spectra, each
with a value of $\Gamma=0.58$, and a normalization on large scales based on 
the COBE four year results (Bunn \& White 1997).  
Each assumes a value of $h=0.65$, and
for comparison we plot both the numerical model and analytical fit
with $\Omega_{0}=1$, $\Omega_{b}=0.057$, the numerical result 
for $\Omega_{0}=0.89$, $\Omega_{b}=0$, 
and the numerical result for $\Omega_{0}=1.37$,  
$\Omega_{b}=0.26$, which has the same value of $\Omega_{b}/\Omega_{0}$ as 
that found in the open model in panel b.  
Note that the $\Omega_{b}=0$ numerical model and the 
analytical fit are almost indistinguishable.  This is because the two have 
identical values of $\Gamma$ and neither breaks this degeneracy with wiggles.
Panel b shows an analytic fit and a numerical model of an $\Omega_{0}=0.3$, 
$\Omega_{b}=0.057$ power spectrum.
Panel c shows the ratio of the analytic fit to the numerical 
model power spectrum for the flat, $\Omega_{b}=0.057$ case.  Finally, panel d
shows the same relationship for the open case.  It is clear from 
panels c and d that the wiggles, while present in the flat universe model, 
have a much greater effect in an open universe.  
We propose to estimate how well $\Omega_{b}$ can be determined by examining
these features.  The emphasis in this paper is not to present a method of
measuring $P(k)$, but to estimate errors on quantities on which $P(k)$
depends, especially $\Omega_b$. 

The outline of the paper is as follows.  In \S~\ref{sec:fisher} we discuss the use of the
Fisher information matrix in determining parameter uncertainties, and 
in \S~\ref{sec:method} we present the details of our
method. \S~\ref{sec:results} contains the
results of that method for several different cosmological models. 
In \S~\ref{sec:complications}, we discuss other complicating factors which enter into the
analysis of redshift survey power spectra.  
Finally, in \S~\ref{sec:conclusions}, we present our conclusions.  In Appendix A, we present 
an alternate expansion of the SDSS density field, which may be useful for 
future computational work.

\section{How Well Can We Determine Parameters?}
\label{sec:fisher}

Suppose that we are able to parameterize a power spectrum at the 
current epoch according
to some set of parameters, ${\bf \Theta}$.
A moderately complete set might include the
following parameters: ${\bf \Theta} =\{ \sigma_{8}, \Omega_{0}, \Omega_{b},
h, n_{prim}\}$, where the latter four have been defined previously, and 
$\sigma_{8}$ is a normalization constant, the rms density fluctuations within
$8h^{-1}$Mpc spheres.  We assume that other parameters, such as $\Lambda$, are 
known exactly throughout this discussion; we set $\Lambda=0$. 

In addition to a model, we  
have some data, $d_{i}, i=1,...,n$ put into a data vector, ${\bf d}$, where $n$ is
the total number of data elements.  Moreover,
we have a method to relate the raw observables 
(e.g. positions and redshifts of galaxies) to ${\bf d}$.
In order to compare ${\bf d}$ to the model, we can associate some likelihood
function $L({\bf \Theta};{\bf d})$, which
we would like to maximize in order to find the ``true'' cosmological
 parameters, ${\bf \Theta}_{0}$, where
\begin{equation}
L=\frac{e^{-{\bf d}^{\dagger}{\bf C}^{-1}{\bf d}/2}}
{(2\pi)^{N/2}{\rm det}({\bf C})^{1/2}} \ .
\label{eq:likelihood}
\end{equation}
This form assumes that $\langle d_{i}\rangle=0$ and that the expected
distribution of ${\bf d}$ is a multivariate Gaussian.

The covariance matrix, ${\bf C}$, is defined as
\begin{equation}
C_{lm}\equiv \langle d_{l} d_{m}^{\ast} \rangle \ .
\end{equation}
  
Following Tegmark, Taylor \& Heavens (1997, hereafter TTH), 
we use the Fisher information matrix to 
determine the minimum possible uncertainties in these parameters.
The Fisher matrix is defined as
\begin{equation}
F_{ij} \equiv
\left< \frac{\partial^{2}\cal{L}}{\partial \theta_{i} \partial \theta_{j}}
\right>\ , 
\end{equation}
where ${\cal L}=-\ln L$.  This may further be reduced to (Vogeley \&
Szalay 1996; TTH):
\begin{equation}
F_{ij}=\frac{1}{2}{\rm Tr}\left[{\bf A}_{i}{\bf A}_{j}\right] \ ,
\label{eq:fisher-form} 
\end{equation}
where the matrices 
${\bf A}_{i}\equiv {\bf C}^{-1}\partial {\bf C}/\partial
\theta_{i}$.

The Fisher matrix allows us to
determine the minimum possible errors for each parameter 
in a reasonable way via the Cram\`er-Rao inequality (see TTH for a discussion).  
Regardless of {\em how} the actual parameters are measured,
one cannot hope to do better than:
\begin{equation}
\Delta \theta_{i}\geq \left({\bf F}^{-1}\right)_{ii}^{1/2}.
\label{eq:Cramer-Rao} 
\end{equation}
In the limit where all parameters but the $i^{th}$ are known with certainty, 
this becomes $\Delta \theta_{i}\geq 1/\sqrt{F_{ii}}$.

\section{Method}
\label{sec:method}

In this section, we discuss how we define our data {\bf d}, and
determine its covariance matrix, {\bf C} (\S~\ref{sec:compute-C}).  
Our aim is to use
equation~(\ref{eq:fisher-form}) to determine the Fisher information
matrix {\bf F} for the SDSS redshift survey power spectrum, and thus
put an upper limit on the expected error in $\Omega_b$ via
equation~(\ref{eq:Cramer-Rao}). 

The SDSS northern redshift survey will cover 
an elliptical cone of $\pi$ steradians to an approximate depth of 
600$h^{-1}$Mpc ($z\simeq 0.2$), with redshifts for approximately 
$10^{6}$ galaxies.

\begin{figure}
\centerline{\psfig{figure=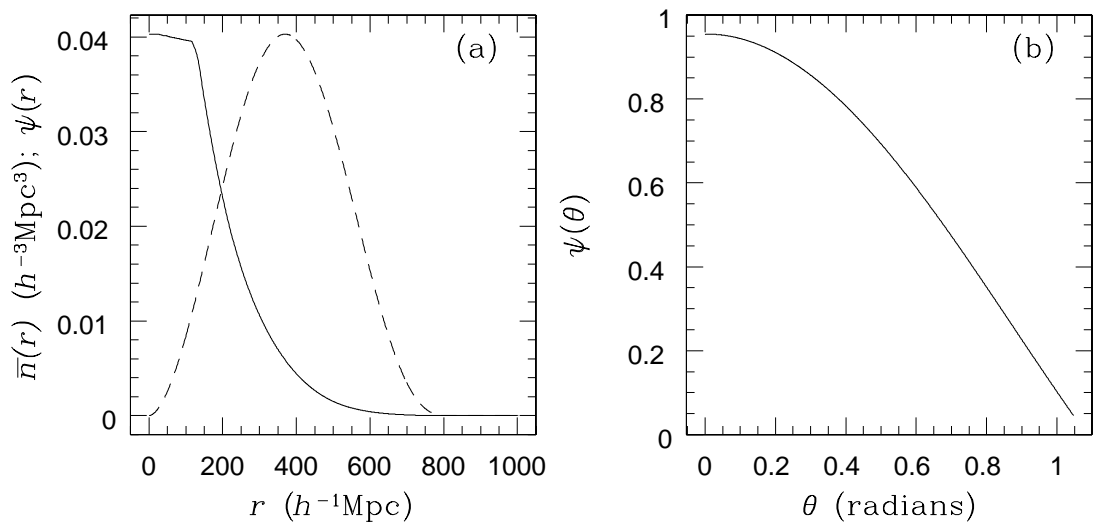,height=6in,angle=0}}
\vspace{-2in}
\caption{(a) The selection function (solid), and the 
radial weight $\psi(r)$ used in weighting the galaxy catalog.
(b) The angular behavior of the 
weight, $\psi(\theta)$, where $\theta$ is the angle from the
central axis of the survey cone.  Since the radial and angular parts of the
weighting function are completely separable, the normalization of each
component is arbitrary.}
\label{fg:sdss}
\end{figure}

D. Weinberg (private 
communication) has created a mock SDSS redshift sample 
based on an $N$-body simulation
by C. Park and J. R. Gott.  We have used this mock survey to calculate a
selection function:
\begin{equation}
\overline{n}(r)=\frac{3}{\omega}\sum_{i,gal}\frac{1}{d^{3}_{max,i}}\ ,
\end{equation}
where $\omega$ is the solid angle of the survey and $d_{max,i}$ is the maximum
distance that galaxy $i$ could be placed
 and still pass the selection criteria of the
survey.  The sum is carried out only over those galaxies for which 
$d_{max,i}>r$.  The selection function for the SDSS is plotted in 
Figure~\ref{fg:sdss}.  We assume that the selection function 
has no angular dependence.

Our primary limitations in the calculation of {\bf F} involve the computation,
storage, and inversion of the covariance matrix, {\bf C}.  All three of these considerations
will be immensely simplified if we can make {\bf C} as close to diagonal
as possible.  We reduce
covariance by an appropriate choice for the data {\bf d}
(\S~\ref{sec:compute-C}) and with a clever choice of weighting
(\S~\ref{sec:psi-choice}).  For the SDSS geometry, 
we find that we can indeed approximate
the covariance matrix as diagonal for scales smaller than $1/8$ the
largest dimension of the survey (\S~\ref{sec:C-size},
\ref{sec:C-cut}).  

\subsection{\em Computing the Covariance Matrix}
\label{sec:compute-C}
We must first specify the form of the dataset, ${\bf d}$, that 
we will use.  The method we use can be described as a 
``poor man's'' Karhunen-Lo\`{e}ve (K-L) transform. The K-L transform 
(Vogeley \& Szalay 1996; TTH) is a method of extracting the 
statistically independent elements of 
a dataset that provide the maximum ratio of signal to noise and hence most
strongly constrain the cosmological parameters.  
Emulating the K-L transforms, we wish to minimize the
values of the off-diagonal elements of the covariance matrix, since we
will ultimately need to invert it (equation~\ref{eq:likelihood}).  If
the survey volume were a cube with a uniform selection function, 
the Fourier modes of the density field with wavelengths along each
axis given by integral
fractions of the cube length, would indeed have no covariance for
non-identical modes; they would be the K-L modes.  We thus embed our survey
volume in a cube of length $L$, which is specified by the largest dimension
of the survey:
\begin{equation}
L=2\ r_{max}\, \sin \theta_{max}\ ,
\end{equation}
where $r_{max} = 809\,h^{-1}$ Mpc is chosen where $\overline{n}(r)$ first
goes to zero, 
and $\theta_{max} = 1.13$ rad is
the largest off-axis angle of the survey. 
This gives $L = 1466 h^{-1}$Mpc.

We use as our dataset {\bf d} the
Fourier transform of the real space density field\footnote{We discuss
redshift-space distortions in \S~\ref{sec:z-distortions}.} with 
\begin{equation}
{\bf k}(n_x,n_y,n_z)=\frac{2\pi n_x}{L}\hat{\bf x}+
\frac{2\pi n_y}{L}\hat{\bf y}+\frac{2\pi n_z}{L}\hat{\bf z}\ ,
\end{equation}
where $n_{x}$, $n_{y}$ and $n_{z}$ are non-negative integers.  Because
the SDSS geometry is of course not cubic, this scheme will not exactly
diagonalize the covariance matrix, but we will see in
\S~\ref{sec:C-cut} that we will be able to make the approximation that
it is diagonal at large $|k|$.  We order the modes in concentric
cubes in {\bf k} space, such that they are roughly ordered in
increasing value of $|k|$\footnote{Note that unlike the K-L modes, we
have not ordered the $d_i$ in decreasing signal-to-noise ratio.}.  We
refer to ${\bf k}_{l}$ as the $l^{th}$ value of ${\bf k}$ in this
ordering.  Thus there is a rough correspondence between increasing $l$
and decreasing scale.  As the $\Omega_b$-induced wiggles are prominent
on large scales, we will determine our Fisher matrix, and
corresponding uncertainties on parameters, as a function of $n$, the
maximum value of $l$ which we probe.

This is clearly not the only basis set which will partially diagonalize
the covariance matrix from the outset.  In Appendix A we discuss a set
of basis vectors constructed from spherical Bessel functions and
the spherical harmonics which will produce very little covariance within the
SDSS geometry.

Following Fisher \etal\ (1993), we estimate the density
field of galaxies as a sum of discrete points:
\begin{equation}
\delta({\bf r})=\sum_{i,gal}\frac{\delta^{D}({\bf r}-{\bf r}_{i})} 
{\overline{n}({\bf r})}-1 \ ,
\label{eq:direct}
\end{equation}
where 
$\overline{n}({\bf r})$ is the selection function,
the expected density of observed galaxies at ${\bf r}$ in the absence of
clustering.  We allow ourselves a weighting function, 
$\psi({\bf r})$,  in order to further reduce covariance and to minimize  shot
noise (see \S~\ref{sec:psi-choice}).
The Fourier Transform of the density field, $\hat\delta ({\bf k})$, is
then taken.  Our Fourier convention here and throughout is:
\begin{equation} 
\hat{f}({\bf k})=\int d^{3}{\bf r}\ e^{i{\bf k}\cdot {\bf r}} f({\bf r})\ ,
\end{equation}
 and
\begin{equation}
f({\bf r})=\frac{1}{(2 \pi)^{3}}\int d^{3}{\bf k} 
\ e^{-i{\bf k}\cdot {\bf r}}\hat{f}({\bf k})\ .
\end{equation}

The Fourier modes of the weighted density are then:
\begin{equation}
\hat{\delta} ({\bf k})=\int d^{3}{\bf r}\  
e^{i{\bf k}\cdot {\bf r}}\psi({\bf r})
\delta({\bf r})\ .
\end{equation}
Thus ${\bf d}$ is the set of $\hat\delta({\bf k}_{l})$, for the ordering
of ${\bf k}_{l}$ described above.
Substituting equation~(\ref{eq:direct}) gives:
\begin{equation}
\hat{\delta}({\bf  k})=
\sum_{i}\frac{\psi({\bf r}_{i})e^{i{\bf k}\cdot{\bf r}_{i}}}
{\overline{n}({\bf r}_i)}-W({\bf k})\ ,
\end{equation}
where the window function, $W({\bf k})$, is defined as:
\begin{equation}
W({\bf k})\equiv\int d^{3}{\bf r}\ \psi({\bf r})e^{i{\bf k}\cdot {\bf r}}\ .
\end{equation}

Taking the product of any two of these modes, we can compute the covariance:
\begin{eqnarray}
\hat{\delta}({\bf k}_{1})\hat{\delta}({\bf k}_{2})^{\ast}=
\left(\sum_{i}\frac{\psi({\bf r}_{i})}{\overline{n}({\bf r}_{i})}
e^{i{\bf k}_{1}\cdot {\bf r}_i}\right)
\left(\sum_{j}\frac{\psi({\bf r}_{j})}{\overline{n}({\bf r}_{j})}
e^{-i{\bf k}_{2}\cdot {\bf r}_{j}}\right) \nonumber \\
-W({\bf k}_{1})\sum_{i}\frac{\psi({\bf r}_{i})}{\overline{n}({\bf r}_{i})}
e^{-i{\bf k}_{2}\cdot {\bf r}_{i}}
-W({\bf k}_{2})^{\ast}\sum_{i}\frac{\psi({\bf r}_{i})}
{\overline{n}({\bf r}_{i})}
e^{i{\bf k}_{1}\cdot {\bf r}_{i}}
+W({\bf k}_{1})W({\bf k}_{2})^{\ast}\ .
\label{eq:covar}
\end{eqnarray}

Following Fisher \etal\ (1993) we take the expectation value of this 
expression and after some algebra recover the expression:
\begin{equation}
\langle \hat{\delta}({\bf k}_{1})\hat{\delta}({\bf k}_{2})^{\ast}\rangle =
\frac{1}{(2\pi)^{3}}\int d^{3}{\bf k}' P({\bf k}') W({\bf k}_{1}-
{\bf k}')W({\bf k}_{2}-{\bf k}')^\ast
+\int d^{3}{\bf r} 
\frac{\psi({\bf r})^{2}e^{i({\bf k}_{1}-{\bf k}_{2})\cdot {\bf r}}}
{\overline{n}({\bf r})}\ .
\label{eq:covmat}
\end{equation}
 
If the weighting function is normalized such that $\int d^3 {\bf r}\  
\psi({\bf r})=1$,  
the diagonal elements of the covariance matrix are:
\begin{equation}
C_{ll}=\tilde{P}({\bf k}_{l})+P_{shot}\ ,
\end{equation}
where 
\begin{equation}
\tilde{P}({\bf k}_{l})=\frac{1}{(2\pi)^{3}}\int d^{3}{\bf k}' P({\bf k}')
|W({\bf k}_{l}-{\bf k}')|^{2},
\label{eq:diag}
\end{equation}
and
\begin{equation}
P_{shot}=\int d^3{\bf r}\frac{\psi({\bf r})^{2}}{\overline{n}({\bf r})}\ ,
\end{equation}
which matches the result expected from traditional methods of power
spectrum estimation (Fisher \etal\ 1993; Feldman, Kaiser, \& Peacock 1994,
hereafter FKP; Park \etal\ 1994).

\subsection{\em The Optimal Choice for $\psi({\bf r})$}
\label{sec:psi-choice}
We now discuss how the weighting function, $\psi({\bf r})$, is chosen.
First, the weighting function must reflect the survey geometry.  That is, 
it must go to zero outside the survey and be well-defined within it.
It must also minimize shot noise by giving 
regions with higher $\overline{n}({\bf r})$ a higher weight.
Finally, we wish to minimize covariance
between modes.  Tegmark (1995) gives a clever solution 
which minimizes covariance in an arbitrary geometry, and we shall use
his algorithm throughout.  The weighting function is the solution
to the Schr\"{o}dinger's Equation:
\begin{equation}
\left[ -\frac{1}{2}\nabla^{2}+\frac{\gamma}{\overline{n}({\bf r})}
\right]\psi({\bf r})=
E\psi({\bf r})\ ,
\label{eq:tegmark}
\end{equation}
where $E$ is the smallest eigenvalue
of the system, and $\gamma$ is a constant with units of length$^{-5}$
which balances the desire to minimize covariance while
maximizing the signal to noise ratio.  We used $\gamma=10^{-8}h^{5}$
Mpc$^{-5} \approx n_{gal}\,L^{-5}$
throughout,
but found similar results for values in the range
$10^{-7}-10^{-9}h^{5}$ Mpc$^{-5}$.  

Since the SDSS has an isotropic selection function within the survey 
footprint, equation~(\ref{eq:tegmark}) is separable; 
we plot the radial and angular parts of the wave-function,
 $\psi({\bf r})$ in Figure~\ref{fg:sdss},
where $\psi({\bf r})\equiv\psi(r)\psi(\theta)$.
The actual footprint of the survey has a small but finite
ellipticity, $e\simeq 0.1$.
Rather than introducing a third, azimuthal, coordinate
in solving Schr\"{o}dinger's equation, we treat the angular shape of
the survey as circular, and then simply stretch the wave-function 
appropriately.

\subsection{\em Isolating the Independent Modes}
\label{sec:independent}

We have made our choice of modes in order to 
minimize the off-diagonal elements in the covariance matrix.  However,
we have not guaranteed that all of our modes are linearly
independent.  If the survey volume is $V$,
then we can imagine gridding the density
distribution of the survey into $N^{3}$
cells.  For large $N$,
we can fully describe the density field on scales of $L/N$ with 
$f_{V}\times N^{3}$ numbers, where:
\begin{equation}
f_{V}\equiv\frac{V}{L^{3}}\ .
\label{eq:f-sub-V} 
\end{equation}
Thus we expect only a fraction, $f_{V}<1$,
of our Fourier modes to be independent on small 
scales.   For the geometry of the SDSS, this fraction is given by 
$f_{V} = 0.176$. 

Since, in general,
this suggests that the covariance matrix is singular, we can use singular
value decomposition (SVD), to create a new basis set.  Examining the 
first $n$ Fourier modes, we 
determine two unitary matrices,
${\bf u}$ and ${\bf v}$ such that
\begin{equation}
{\bf C}={\bf u}\ {\bf \cal C}\  {\bf v}^{\dagger}\ ,
\end{equation}
where ${\cal C}$ is 
a diagonal $n\times n$ matrix containing the eigenvalues, $s_{i}$, 
of ${\bf C}$ in decreasing order.  
Thus, by truncating at some critical value of 
$s_{m}/s_{1}$, we may look at only the first $f_{s}=m/n$ 
modes.  For very large values of $n$,
it is our expectation that $f_{s}$ approximately approaches  $f_{V}$, since both
represent a quantitative account of the amount of information per
Fourier mode.  In general, however, $f_s < f_V$ at large $n$, 
because ostensibly
independent modes may be of  
such low signal to noise ratio that SVD rejects them. 

We label ${\bf \cal C}_{m}$ as the upper $m\times m$
elements of ${\bf \cal C}$, as all other elements are 
very close to zero.  We define the $n\times m$ matrices
${\bf u}_{m}$ and ${\bf v}_{m}$ as the
first $m$ column vectors of ${\bf u}$ and ${\bf v}$, respectively. 
These matrices can be used to transform 
an $n\times n$ matrix into an $m\times m$ matrix.  
We can similarly transform
any arbitrary matrix, ${\bf D}$ to the basis set by:
\begin{equation}
{\bf \cal D}_{m}={\bf u}^{\dagger}_{m}\ {\bf D}\ {\bf v}_{m}\ .
\end{equation}

By transforming $\partial {\bf C}/\partial\theta_{i}$ 
in this way, we can readily compute the
matrices ${\bf A}_{i}={\bf \cal C}^{-1}_{m}{\bf \cal C}_{m,i}$, where 
${\bf \cal C}_{m,i}$
is the transform of $\partial {\bf C}/\partial \theta_{i}$, and 
$({\cal C}_{m}^{-1})_{ab}=\delta^{K}_{ab}/s_{a}$, from which
follows the Fisher matrix.  

\begin{figure}
\centerline{\psfig{figure=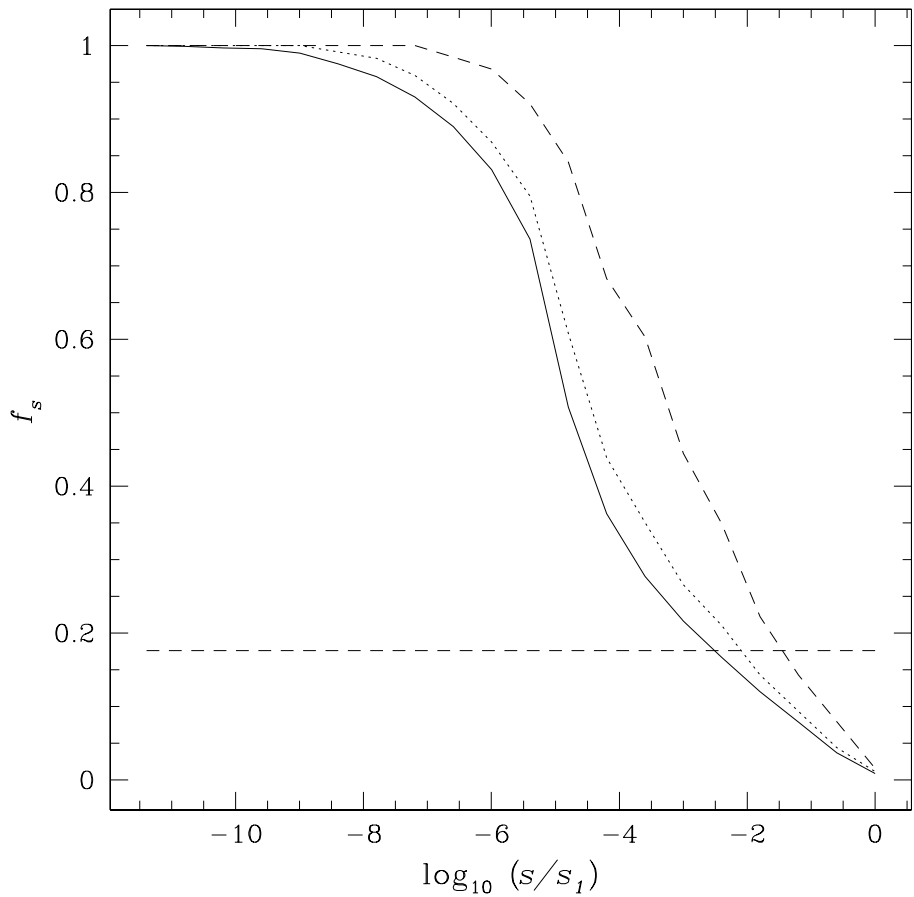,height=5in,angle=0}}
\caption{  A plot of the fraction of eigenvalues $s$ that are greater
than some fractional value, for several different 
maximum numbers of modes.  The dotted line is 63 modes, 
the dashed line is 342 modes, and the solid line is 
999 modes. The horizontal line represents
$f_{V}=0.176$.  
As $n$ increases, the curves asymptote to a uniform form, and that at $n=999$ 
modes, a value of $s_{m}/s_{1}=0.004$ gives approximately $f_{s}=f_{V}$.}
\label{fg:modesize}
\end{figure}

In order to determine the condition number $s_m/s_1$ at which $f_s$
approaches $f_V$, we look at the distribution function of eigenvalues
of the covariance matrix. 
Figure~\ref{fg:modesize} shows this  distribution for 
covariance matrix sizes $n = 63, 342$, and 999, corresponding to $k =
(3, 6, 9)\times 2\,\pi/L$.  This was done
assuming $P(k) = \rm constant$ and ignoring shot noise, 
because we want to 
probe the geometry of the survey, and not a particular power spectrum model.
The horizontal dashed line is drawn at $f_V = 0.176$, which crosses
the $n = 999$ line at $s_m/s_1 = 0.004$.  For $n > 342$, the
distribution function asymptotes to a uniform curve, and therefore we
can choose the condition number to be 0.004 independent of our value
of $n$, and match the requirement $f_s \simeq f_V$. 

\subsection{\em The Size of the Covariance Matrix}
\label{sec:C-size}

Calculating the Fisher matrix requires performing an SVD on
the covariance matrix.  If we make no simplifying assumptions, 
a grid of $N$ different values of ${\bf k}$ 
requires an $N \times N$ matrix.  Performing the SVD requires $\simeq N^{2}$
calculations, and transforming a matrix to the basis set requires 
$\simeq N^{3}$ calculations, not to mention the fact that 
each element in the matrix is a three dimensional integral over a non-symmetric
function (equation~\ref{eq:covmat})!  For a survey of the size of the SDSS 
($\sim 10^{6}$ particles), it  is reasonable to have as many independent 
modes, and using the method outlined
above, vastly overwhelm the limits of both computing time and of storage.

However, our task may be simplified somewhat.   
We have reason to suspect that that after some critical mode, $n_{crit}$,
we may approximate the covariance matrix as being approximately diagonal, since
at small scales, the convolution in equation~(\ref{eq:diag}) becomes more
and more an unbiased estimator of $P(k)$, as the edge effects become
less and less important. 

\begin{figure}
\centerline{\psfig{figure=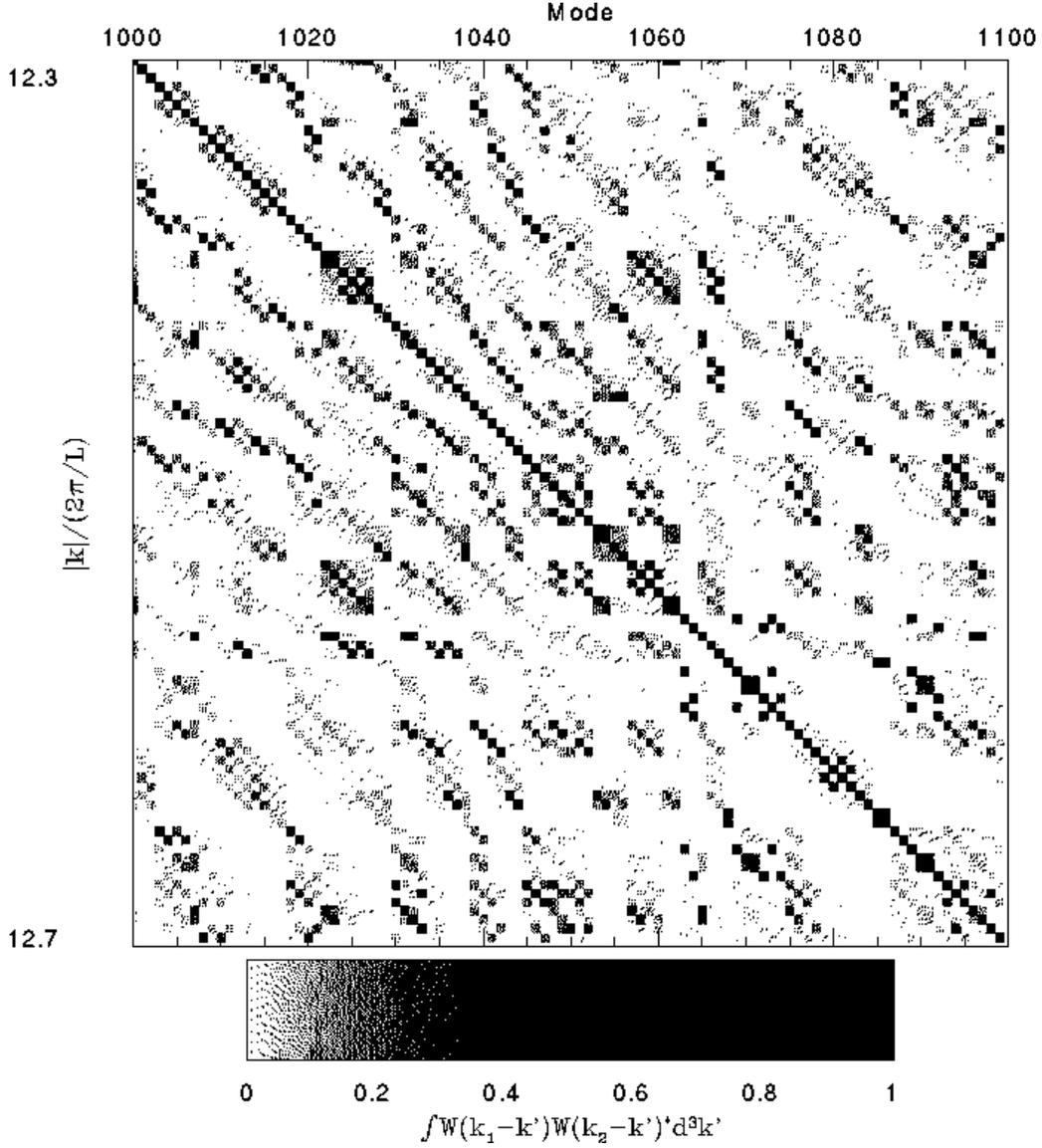,height=6in,angle=0}}
\caption{A grey-scale map of the weighting used in the covariance matrices,
ordered in increasing
value of $|k|$.  The weighting goes as: $\int d^{3}{\bf k}'
W({\bf k}'-{\bf k}_{1}) W({\bf k}'-{\bf k}_{2})^{\ast}$.  (a) 1/100
 of the first
$10^{4}$ modes. (b) The first 100 modes. 
(c)The $1000-1100^{th}$ modes.  Note that the matrix gets sparser and
sparser as we look at higher and higher modes.}
\label{fg:quilt}
\end{figure}

Figure~\ref{fg:quilt} represents the increasing sparseness of the covariance
matrix pictorially, where we assume the SDSS geometry and selection
function, again using the simplified model of a constant power spectrum 
with no shot noise.
In Figure~\ref{fg:quilt}a, we show the first 10000
modes of the covariance matrix (selecting each $100^{th}$ mode).  
Figure~\ref{fg:quilt}b shows the covariance matrix for the first 100 terms, 
and Figure~\ref{fg:quilt}c shows
the $1000^{th}$ to $1100^{th}$ terms.  In each case, the shading is linearly
proportional to the covariance between the modes.  The off-diagonal
terms become smaller for the higher modes as the edge effects become
less important. 
We should therefore be
able to truncate analysis of the full covariance matrix at some point,
as we discuss in the following subsection. 

However, we cannot simply consider the covariance matrix to be completely
diagonal beyond some $n_{crit}$, given the argument in
\S~\ref{sec:independent} about the overcounting of independent modes. We therefore consider a 
likelihood function for the $n_{diag}$ modes from $n_{crit}+1$ to
$n_{crit}+n_{diag}$, using the diagonal covariance matrix approximation:
\begin{equation}
{\cal L}^{diag}=
\frac{f_{V}}{2}\sum_{n=n_{crit}+1}^{n_{crit}+n_{diag}}
\left(\frac{|d_{n}|^{2}}{C_{nn}}+\ln C_{nn}\right)+
\frac{f_{V}}{2}\,n_{diag}\, \ln (2\pi)\ .
\end{equation}

From this, the diagonal estimate Fisher matrix
elements are:
\begin{equation}
F^{diag}_{ij}=\frac{f_{V}}{2}\sum_{n=n_{crit}+1}^{n_{crit}+n_{diag}}
\frac{\partial C_{nn}/\partial\theta_{i}}{C_{nn}}
\frac{\partial C_{nn}/\partial\theta_{j}}{C_{nn}}\ .
\end{equation}
 
Thus, we will treat the covariance matrix in two parts: a square
matrix {\bf C}$^{square}$ with $k<k_{n_{crit}}$, and a diagonal matrix
${\bf C}^{diag}$ with $k\geq k_{n_{crit}}$.  Simple algebra
demonstrates ${\bf F}={\bf F}^{square}+{\bf F}^{diag}$.  We now turn
to the determination of an appropriate value for $n_{crit}$. 

\subsection{\em At What Point May We Use the Diagonal Covariance Matrix?}
\label{sec:C-cut}

In order to determine at what $n_{crit}$ we can start treating the
covariance matrix as diagonal, we have computed both the full and
diagonal covariance matrices for a power spectrum model with
$\Omega_{0}=1$, $\Omega_{b}=0.057$, and $\sigma_{8}=1.6$, with the
effects of shot noise included.  We performed an SVD on the full
covariance matrix using $s_{m}/s_{1}=0.004$ determined above.  We then
computed the Fisher matrix element corresponding to the normalization
of the power spectrum for both the full and diagonal estimates, as a function of
maximum mode number, $n$.

In order to compare these two estimates, 
let us define the Fisher function $F_{ii}(n)$ as the value of $F_{ii}$
computed for the first $n$ terms of the covariance matrix.
To compare $F_{ii}$ computed
using the full and diagonal covariance matrices,
we introduce an improvement function:
\begin{equation}
{\cal I}(n)=\frac{{F}_{ii}^{diag}(n)}
{F_{ii}^{full}(n)}\ .
\end{equation}
 It is our hope, of course, that for some value of $n$,
the improvement function
tends to unity, after which we can safely apply only diagonal terms without a
significant loss of information. 

\begin{figure}
\centerline{\psfig{figure=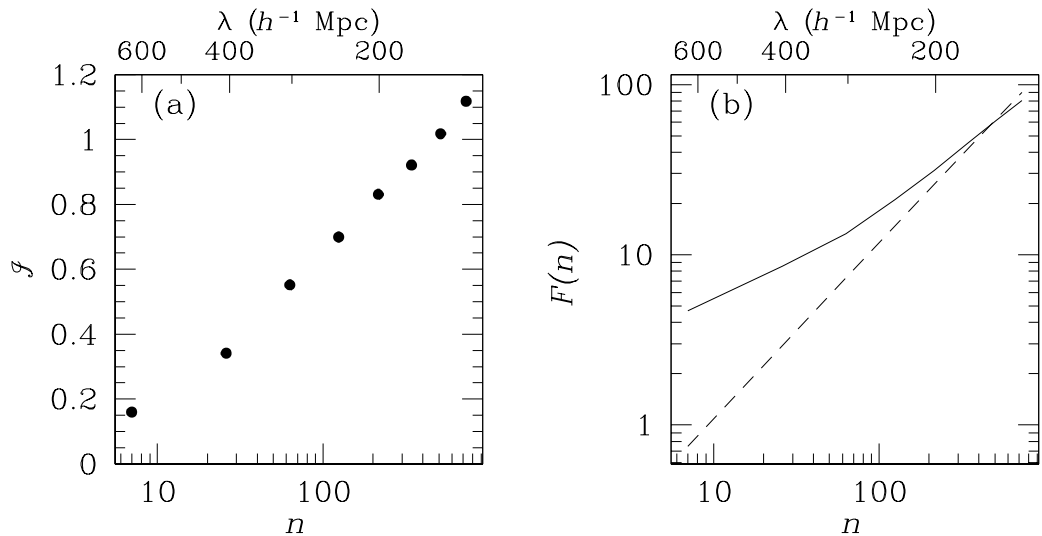,height=6in,angle=0}}
\vspace{-2in}
\caption{(a) The improvement function, ${\cal I}(n)$, for the
normalization of the power spectrum 
as a function of the number of modes, $n$.  (b) The Fisher function, 
$F(n)$.  The solid line is computed using the full covariance matrix, 
while the dashed line is computed using the diagonal elements only.  
The above analysis goes out to 
$k=8\times 2\pi/L$, which corresponds to the first $728$ modes.}
\label{fg:lim1}
\end{figure}

In Figure~\ref{fg:lim1}, we present the results of this exercise.  
Figure~\ref{fg:lim1}a shows the improvement function as a function 
of mode number, while Figure~\ref{fg:lim1}b shows the 
behavior of the Fisher function, $F(n)$, 
for the full matrix (solid)
and the diagonal elements only (dashed). 
The two curves approach one another at $n=728$. 

We thus adopt $n_{crit} = 728$, and treat the covariance matrix as
diagonal for larger $l$.  In summary, we have several reasons to argue
that we can do this:
\begin{enumerate} 
\item The covariance matrix becomes increasingly sparse at higher values of 
$|k|$ (Figure~\ref{fg:quilt});
\item The Fisher information matrix using the diagonal and the full
covariance matrix gives similar results at $n = 728$ (Figure~\ref{fg:lim1});
\item By $n = 728$, the distribution function of eigenvalues
(Figure~\ref{fg:modesize}), and
therefore the condition number required to make $f_{s}\simeq f_{V}$,
asymptotes to a constant value. 
\end{enumerate}

\section{Results}
\label{sec:results}

\subsection{\em Expected $\Omega_{b}$ Uncertainties for Various Cosmologies}

In this section, we compute the  uncertainties 
of cosmological parameters used in a typical model power spectrum using
the Fisher information matrix.  We will present 3 models, each with $h=0.65$,
$n_{prim}=1$, and $\Lambda=0$.  
Model 1 is a flat universe
($\Omega_{0}=1$) with a baryonic component consistent with nucleosynthesis
estimates ($\Omega_{b}=0.024 h^{-2}$) and a value of 
$\sigma_{8}=1.6$.  Model 2 is a flat universe 
with no baryonic component and $\sigma_{8}=1.6$.  
Finally, Model 3 is an open universe,
$\Omega_{0}=0.3$, with $\Omega_{b}=0.024 h^{-2}$ and $\sigma_{8}=0.64$. 
All three models are consistent with COBE normalization for
their respective cosmologies (Bunn \& White 1997 and references therein).
The power spectra of Models 1 and 3 were shown in Figure~\ref{fg:power}.

We also include a Bayesian prior Fisher matrix, which takes into
account the fact that we {\em do} have prior information about $h$,
$n_{prim}$, and $\Omega_{0}$.  We assume that these quantities have
Gaussian-distributed, uncorrelated (and quite conservative) errors
$\Delta \Omega_{0}=3$, $\Delta h=0.5$, and $\Delta n_{prim}=1$.  
The prior log-likelihood function is:
\begin{equation}
{\cal L}^{prior}({\bf \Theta})=\sum_{i}\left(
\frac{(\theta_{i}-\theta_{i,0})^{2}}{\Delta\theta_{i,0}^{2}}
+\ln (\Delta \theta_{i,0})\right)
+N\, \ln (2\pi)\ .
\end{equation}
The contribution to the Fisher matrix is:
\begin{equation}
F_{ij}^{prior}=\Delta\theta_{i,0}^{-2}\delta^{K}_{ij}\ ,
\end{equation}
where $F^{total}_{ij}=F^{SDSS}_{ij}+F^{prior}_{ij}$.  

Note 
that as we add data, the measured uncertainties will quickly dominate, and
hence our final error estimates will not be
strongly linked to our Bayesian priors.

\begin{figure}
\centerline{\psfig{figure=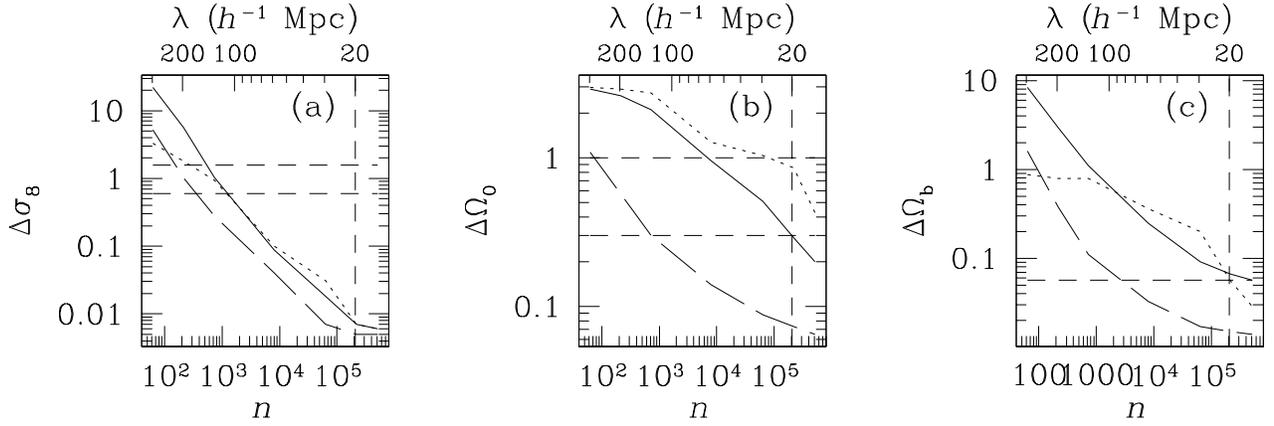,height=7in,angle=0}}
\vspace{-1in}
\caption{The computed uncertainties for (a) $\sigma_{8}$, (b) $\Omega_{0}$, and 
(c) $\Omega_{b}$ for various cosmological models.  The solid line is Model 1,
the flat, $\Omega_{b}=0.057$ model.  The dotted line is Model 2, the
flat $\Omega_{b}=0$ model.  The long dashed line is Model 3, 
the $\Omega_{0}=0.3$, $\Omega_{b}=0.057$ model.  The short-dashed
vertical line is at the mode corresponding to $\lambda=20h^{-1}$Mpc, the
approximate limit of the linear regimes.  The horizontal dashed lines are the 
actual values used in each of the models.}
\label{fg:error1}
\end{figure}

Figure~\ref{fg:error1} 
shows $\Delta \theta_{i}=(F^{-1})_{ii}^{1/2}$, the computed parameter
uncertainties for each of these models.  
If we include all the Fourier modes
down to a physical scale of $20h^{-1}$Mpc (where the evolution of the 
power spectrum is expected to become non-linear), we find that the uncertainty
in $\Omega_{b}$ is $0.070$, $0.060$, and $0.016$ for the three models, 
respectively.
Models 1 and 3 have a value of $\Omega_{b}=0.057$, which means
that, at the 
very least, we could check for inconsistency with $\Omega_{b}=0$ at
the 0.8 and 3.5 $\sigma$ level, respectively.  This suggests
that studying redshift surveys in 
an open universe will give us a much better handle on $\Omega_{b}$
than will a closed universe, as the size of the wiggles depends on
$\Omega_b/\Omega_0$ (Figure~\ref{fg:power}).  

Note that $\Delta \sigma_{8}$ varies by a factor
of $\simeq 2$ from the open to closed models, roughly the ratio of the values
of $\sigma_{8}$ used in each model.  This gives a fractional uncertainty which 
is roughly constant between models. 
Likewise, we estimate similar fractional uncertainties in $\Omega_{0}$ 
for all three models.  

Finally, it worth noting that there is no discontinuity
in the behavior of any of the estimated parameter uncertainties
at $n= 728$ modes, where we start treating ${\bf C}$ as diagonal.
This reassures us that the transition to a diagonal covariance matrix
does not introduce any gross feature in the Fisher matrix. 

\subsection{\em A Different Set of Parameters}

The parameters used in the parameter estimation, $\bf \Theta$, are not
independent of each other; for example, $\Omega_0$, $h$, and $\Omega_b$ are
coupled through $\Gamma$ (equation~\ref{eq:Gamma}).  This is reflected
in the off-diagonal elements of the Fisher matrix, which is large for
pairs that are coupled.  Thus the covariance between
$\Omega_0$ and $\Omega_b$,
\begin{equation} 
\mu_{ij} \equiv {{\left(F^{-1}\right)_{ij}} \over
{\left[\left(F^{-1}\right)_{ii}\left(F^{-1}\right)_{jj}\right]^{1/2}}}
\ ,
\label{eq:mixing-angle} 
\end{equation}
is 30\% on small scales, and appreciably larger on larger scales.  We
thus are motivated to use a set of parameters that are closer to
orthogonal, to minimize their uncertainty. 
In particular, the parameter $\Gamma$ sets the overall shape of the
power spectrum, 
and the amplitude of the 
wiggles goes roughly as $\Omega_{b}/\Omega_{0}$.
 Therefore, we now consider 
the parameter set, ${\bf T}=\{ \sigma_{8}, \Gamma, \Omega_{b}/\Omega_{0},
h, n_{prim} \}$, since these quantities should be closer to independent. 

We use the chain rule to convert between one data set and another:
\begin{equation}
F_{ab}=\left< \frac{\partial^{2}{\cal L}}{\partial T_{a} \partial T_{b}}
\right>
=\sum_{i,j} F_{ij}
\frac{\partial \theta_{i}}{\partial T_{a}} 
\frac{\partial \theta_{j}}{\partial T_{b}}\ .
\end{equation}

\begin{figure}
\centerline{\psfig{figure=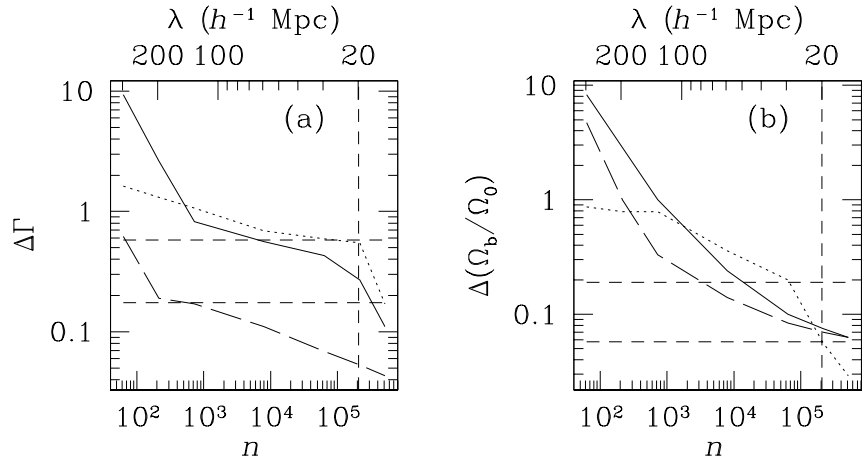,height=5in,angle=0}}
\caption{The computed uncertainties of (a) $\Gamma$ and (b) 
$\Omega_{b}/\Omega_{0}$.
The solid line is Model 1, the dotted line is Model 2, and the long
dashed line is Model 3. }
\label{fg:error2}
\end{figure}

Figure~\ref{fg:error2} plots the results of the parameter rotation.
The uncertainties in $\Gamma$ are about three times larger in the flat 
universe models than the open universe model, corresponding to very
similar fractional errors in each model.  The uncertainties of the parameter
$\Omega_{b}/\Omega_{0}$ behave consistently with the results determined in the
previous section.  That is, all three models produce roughly the same 
uncertainty in $\Omega_{b}/\Omega_{0}$, about $0.06$, corresponding to 
the values of $\Omega_{b}/\Omega_{0}$ inconsistent with zero at the
$\simeq 1\sigma$ level in Model 1, and the $\simeq 3\sigma$ level
in Model 3.  Note that models 1 and 2 have sharp downturns in $\Delta \Gamma$
at around the threshold of the nonlinear 
regime.  This is due primarily to the fact that the
wiggles appear most prominently in flat cosmologies at 
around those scales
(see Figure~\ref{fg:power}).
The covariance between 
$\Omega_{b}/\Omega_{0}$ and $\Gamma$ is only 15\% in this regime,
and thus both parameters can be determined with greater accuracy.

\subsection{\em A Comparison with Band Estimates of the Power Spectrum}

We check for consistency of the above method with another,
perhaps more intuitive estimate of the power spectrum. In this section,
we divide $k$-space into bands, and determine the errors of $P(k)$ in each
band.  From there, we can compute the uncertainties in the power spectrum
parameters using the Fisher formalism.

FKP give an estimate for the uncertainty in P($k$) for an approximately 
isotropic survey.  Though this is not strictly true in our case, this
approach is merely to serve as an order of magnitude consistency check 
on the final errors.  Moreover, on smaller scales, isotropy is a fairly 
good assumption.  The expected error for each band is:

\begin{equation}
\frac{\langle \Delta P(k)\rangle }{P(k)}=
\left({\frac{(2\pi)^{3}\int d^{3}{\bf r}
\ \overline{n}^{4}({\bf r})\psi^{4}({\bf r})
\left[1+\frac{1}{\overline{n}({\bf r})P(k)}\right]^{2}}
{V_{k}
\left[\int d^{3}{\bf r}\ \overline{n}^{2}({\bf r})
\psi^{2}({\bf r})\right]^{2}}}
\right)^{1/2}\ ,
\label{eq:delta-P} 
\end{equation}
where $V_{k}$ is the volume of the shell in $k$-space; we will use the
weight function $\psi$ shown in Figure~\ref{fg:sdss}.  If we take spherical
shells, the volume of the $i^{th}$ shell is
$V_{k}=\frac{4}{3}\pi(k_{i}^{3}-k^{3}_{i-1})$.  We set the bands to be spaced
at increments of $k=2\pi/L$, where $L$ is the characteristic length defined 
in \S~\ref{sec:compute-C}, as the modes should be approximately
independent with this spacing. The FKP model
ignores covariance between modes, and we will use this approximation
here as well. 

\begin{figure}
\centerline{\psfig{figure=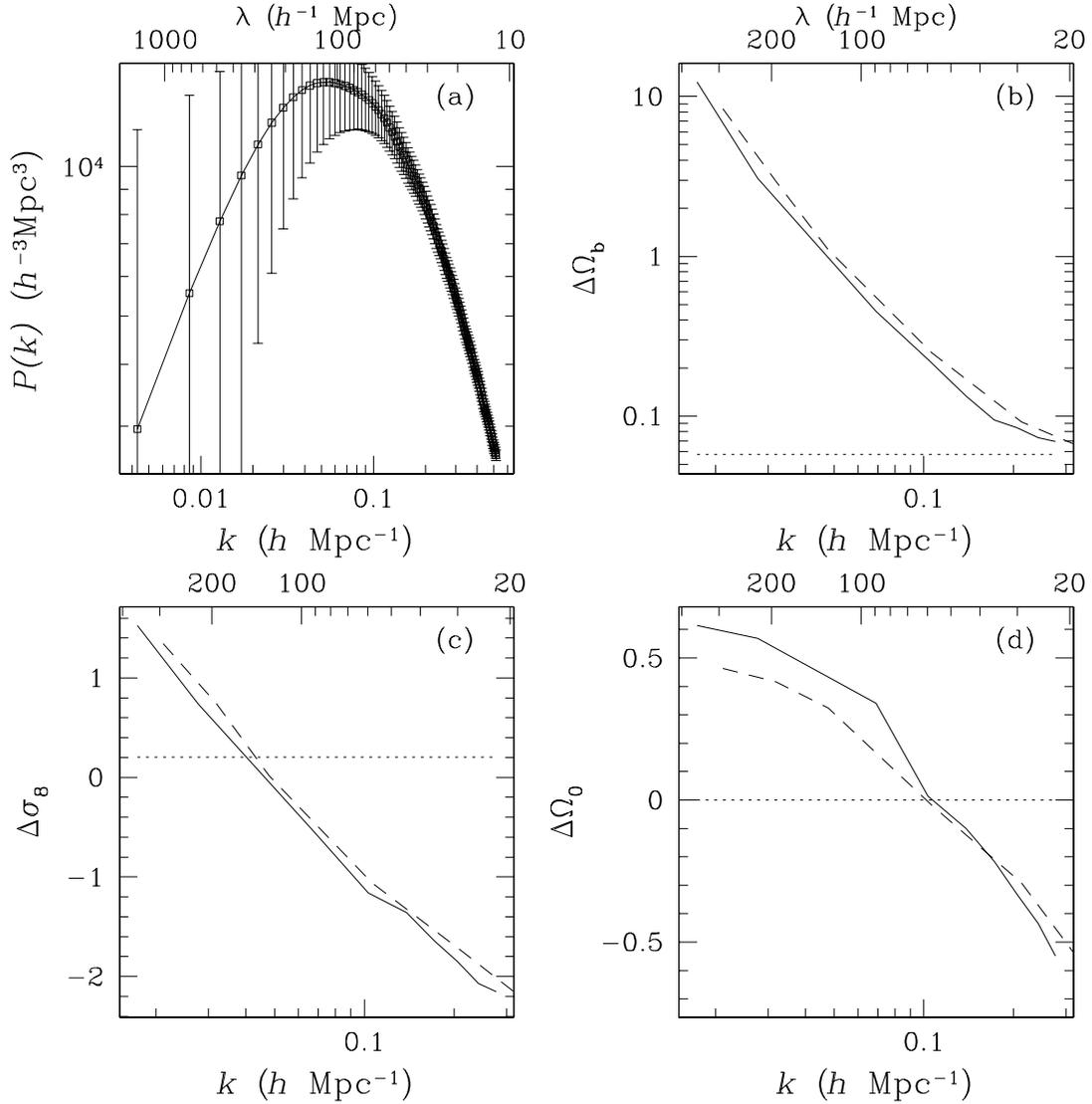,height=6in,angle=0}}
\caption{The associated errors on the Model 1 ($\Omega_{0}=1$,
$\Omega_{b}= 0.057$)
power spectrum and its parameters. (a) The errors on the banded power
spectrum using the FKP error bars.
(b) The uncertainty on $\Omega_{b}$ using FKP errors (solid)
compared with the uncertainty of $\Omega_{b}$ computed using the 
covariance matrix (dashed), as a function of $|k|$.
(c) The errors for $\sigma_{8}$.
(d) The errors for $\Omega_{0}$.}
\label{fg:band}
\end{figure}

The Model 1 power spectrum ($\Omega_{0}=1$, $\Omega_{b}=0.057$) and
the resulting errors are plotted in Figure~\ref{fg:band}. 

The log-likelihood function for some dataset of band-averaged power spectrum 
modes can be determined using a standard $\chi^{2}$ expression because
we have neglected covariance:
\begin{equation}
{\cal L}^{FKP}=\frac{1}{2}\sum_{l}\left(\
\frac{(P_{l}-\tilde{P}_{l})^{2}}{(\Delta P_{l})^{2}}+
\ln (\Delta P_{l})\right)
+\frac{N}{2}\,\ln (2 \pi)\ .
\end{equation}

Here, $P_{l}$ is the value of $P(k_{l})$ computed using a model,
$\tilde P_{l}$ is the measured value of the power spectrum within the
$l^{th}$ band, and $\Delta P_{l}$ is the associated error computed above.
The Fisher information matrix follows straightforwardly (see also Tegmark 1997):
\begin{equation}
F_{ij}=\sum_{l}\frac{\partial P_{l}}{\partial \theta_{i}}
\frac{\partial P_{l}}{\partial \theta_{j}}
\frac{1}{(\Delta P_{l})^{2}}\ .
\label{eq:fisher-fkp} 
\end{equation}

Note that $F_{ij}$ is insensitive to bin size; if we were to take
smaller bins, we would have more terms contributing to the sum in
equation~(\ref{eq:fisher-fkp}), each with smaller $V_k$ and therefore
larger $\Delta P$ (equation~\ref{eq:delta-P}). 
In Figure~\ref{fg:band}, we present the comparison of the final errors on 
$\Omega_{b}$, $\sigma_{8}$, and $\Omega_{0}$
determined in this way with those determined
using the full covariance matrix, as a function of the maximum value of 
$|k|$ used in the analysis.  The two are in fairly good agreement, but in 
general, the band estimate produces smaller estimated uncertainties than
does the covariance matrix analysis.  This is not surprising,
since the band estimates assume that each band can be measured
independently from one another, and thus, it neglects covariance {\em between}
different bands.  This serves to underestimate the true error.
Furthermore, the FKP approximation uses the assumption of isotropy and 
weighs all of the modes {\em within} a band equally
(see Tegmark 1995).  
Finally, in the full covariance matrix analysis, covariance between individual 
modes within a given band can give
information about $P(k)$.  However, both the FKP approximation
and the diagonal covariance matrix analysis ignore this.

\section{Other Complicating Factors}
\label{sec:complications}

We have assumed that the observed galaxy field
reflects the mass density field on the length scales that we probe.
Furthermore, 
we assumed that the redshifts measured represent the actual distances
of the galaxies, and thus we have an accurate, three-dimensional map of
the density field.  We also assumed that the clustering of our density
field behaves in the same way throughout the volume of the survey.  
Finally, we assumed that linear theory describes the evolution of the
power spectrum perfectly down to some scale.  We 
briefly address the validity of these assumptions, 
and what may be done to compensate for them when they do not hold.  A more
comprehensive review of some of these issues appears in Strauss \& Willick (1995).
 
\subsection{\em Evolution of Clustering}

The SDSS survey will
extend to a redshift of $z\simeq 0.2$.  As a result, we would expect that
local structure will have a clustering amplitude which is larger than at the furthest
reaches of the survey due to the growth of structure with time.  Let $A(z)$ be the ratio of power at redshift,
$z$, to redshift $0$; $A(z)=1/(1+z)^{2}$ for $\Omega_{0}=1$, in 
linear theory.
We can correct the density
perturbations at any given redshift, $z$, to account for the cluster evolution
that will take place between $z$ and $z=0$:
\begin{equation}
\hat\delta({\bf k})=\int d^{3}{\bf r}\ e^{i{\bf k}\cdot {\bf r}}
A(z)^{-1/2}\psi({\bf r})\delta({\bf r})\ ,
\end{equation}
and where $z$ and $r$ are related through the standard cosmological
equations. 

Following this through, we find that the window function is redefined as:
\begin{equation}
W({\bf k})=\int d^{3}{\bf r}\ e^{i{\bf k}\cdot {\bf r}}
A(z)^{-1/2}\psi({\bf r})\ .
\end{equation}

Using the revised definition of the window function, the covariance matrix
becomes
\begin{equation}
C_{lm}=\langle \hat{\delta}_{l}\hat{\delta}_{m}\rangle =
\frac{1}{(2\pi)^{3}}\int d^{3}{\bf k}' P({\bf k}') W({\bf k}_{l}-
{\bf k}')W({\bf k}_{m}-{\bf k}')^\ast
+\int d^{3}{\bf r}A(z)^{-1}
\frac{\psi({\bf r})^{2}e^{i({\bf k}_{l}-{\bf k}_{m})\cdot {\bf r}}}
{\overline{n}({\bf r})}\ .
\end{equation}
$A(z)$ depends on $\Omega_{0}$ and to a lesser degree $\Lambda$.  Therefore,
including this effect breaks the degeneracy between $\Omega_{0}$ and $h$,
although further experiments need to be done to quantify how large an
effect this actually is. 

\subsection{\em Redshift Space Distortions}
\label{sec:z-distortions}

Throughout this paper, we have assumed that we know the full three-dimensional
position of each galaxy.  However, in a real survey, we do not know the 
distance of each galaxy, only its redshift, which differ due to peculiar 
velocities.   On small scales, the redshift
distribution differs from the real space distribution via the ``fingers
of God'', the elongated structures seen along the line of sight due to 
motions of galaxies in a virialized system.  On larger scales, gravitational
collapse causes structure to appear compressed along the line of sight. 

If our galaxy survey were confined to a sufficiently small angle on 
the sky, we would be able to use a fairly simple approximation in
linear theory to account for these effects (Kaiser 1987; Cole, Fisher,
\& Weinberg 1994):

\begin{equation}
\hat\delta({\bf k})_{S}=\hat\delta({\bf k})_{R}(1+\beta \mu^{2})\ ,
\end{equation}
where $\hat\delta_{S}$ and $\hat\delta_{R}$ are the Fourier transforms
of the redshift and real density perturbations, respectively, 
$\beta\equiv f(\Omega_{0})/b$, where $b$ is the biasing parameter, 
and $f(\Omega_{0})$ is the logarithmic
derivative of the fluctuation growth rate, and the cosine of the 
angle between ${\bf k}$ and the 
line of sight is $\mu$.  

The more general case is discussed by Zaroubi \& Hoffman (1996).  Using 
linear perturbation theory, they find the general relation between the
redshifted and real Fourier modes.  They show that different values of
{\bf k} become coupled due to redshift distortions.
Since we actually ``measure'' $\hat\delta({\bf k})_{S}$, but the real
density perturbations are given by $\hat\delta({\bf k})_{R}$, one needs
to compute the covariance matrix of the redshift space modes in terms of 
the real space perturbations and their redshift distortions.  This of
course introduces a new parameter, $\beta$, to be included in the
Fisher matrix.  The expressions of Zaroubi \& Hoffman (1996) are quite
complicated, and introduce further coupling between modes, which may
affect the applicability of the techniques used in this paper.  In the
Appendix, we discuss an alternate basis set that has the potential to
reduce this problem for redshift space distortions. 

  Finally, Nakamura \etal\ (1997) and de Laix \& Starkman (1997)
discuss the effects of space curvature on the redshift space
distortions in the SDSS galaxy survey, and show that these effects
need to be taken into account for a proper treatment of redshift space
distortions. 

\subsection{\em Biasing and Selection Effects}

We have further ignored the fact that we are dealing with
the galaxy density field, rather than the unbiased matter density field.
Here we mention several effects that biasing may give. 

First, we have assumed that a galaxy density fluctuation of a certain scale, 
$\delta_{g}({\bf r})$ is equal to the density fluctuation of matter,
$\delta_{DM}({\bf r})$.  Indeed, the two may be related by some constant
of proportionality, $b$, such that 
$\delta_{gal}({\bf r})=b\ \delta_{DM}({\bf r})$.  The normalization parameter,
$\sigma_{8}$, which is proportional to $P(k)^{1/2}$, is also proportional to 
$b$, and thus is not a major concern.  

A more serious issue comes from the fact that biasing is not
necessarily linear, nor is it deterministic.  The bias can be a strong
function of scale and can have finite scatter (cf., Cen \& Ostriker
1992), and hence, the normalization (and the other relevant
parameters) are no longer corrected in a non-trivial way.

We have further made the assumption that the bias is independent of
galaxy luminosity.  Since the galaxies at high redshift are at the
bright end of the luminosity distribution in a flux-limited sample,
the measured galaxy power spectrum at high $z$ will be different from
that found locally if this assumption breaks down. 
We could avoid this problem by using a volume-limited
survey.  However, this would increase the shot noise; further
calculations are needed of the Fisher matrix in this case to see how
this affects $\Delta \Omega_b$. 

Finally, we have made the assumption that there is 
no substantial evolution in galaxy populations since $z\simeq 0.2$.  This 
further complicates matters in that the bias parameter can be a function
of $z$ (e.g., Fry 1996) as well as local density.

These objections notwithstanding, future improvement in our understanding
of evolution, biasing, and the universal luminosity function can certainly
be incorporated into the Fisher formalism.
Once these effects are included in the covariance matrix, the
error estimation proceeds exactly as we have seen, 
with the possible addition of further parameters which describe these
effects. 

\subsection{Nonlinear Effects}

We have treated the evolution of the power spectrum as well described 
by linear theory down to
a scale of $20 h^{-1}$Mpc, and assumed that everything below that
scale was nonlinear.  However, this may not be a concrete limit.  Both
analytic (Taylor \& Hamilton 1996; Jain \& Bertschinger 1994) 
and numerical (Baugh \& Efstathiou
1994) analyses suggest that for some models, the linear regime may
extend to even smaller scales.  This produces two competing effects.
First, nonlinearity tends to smooth out features in the power
spectrum, reducing the size of the wiggles, and thus, limiting the
information about $\Omega_{b}$.  However, if we have a good model of
the power spectrum which does extend well into the non-linear regime,
we can also extend our covariance matrix to smaller scales.  Though we 
expect features like the wiggles to be smoothed out (cf., Tegmark 1997), we
nevertheless may expect to get more information on $\Gamma$, and with that
parameter more tightly constrained, our uncertainties in $\Omega_{b}$ decrease
as well due to the covariance between them. 

\section{Conclusions}
\label{sec:conclusions}

We have discussed the possibility of measuring $\Omega_{b}$ from the 
power spectrum in the forthcoming SDSS.  In order to simplify matters, we
have assumed uniform 
biasing, and that there are no redshift space distortions and no
evolution of either the density fields or of the constituent
galaxies.  Our approach is to use the Fisher information matrix, which
uses the covariance matrix of the data to put minimum error bars on
derived parameters via the Cram\`er-Rao inequality. 

We decompose  the density field of a galaxy redshift survey into 
discrete Fourier modes in units of $2\pi /L$, where the characteristic
length, $L = 1466h^{-1}$ Mpc, is given by the survey geometry.  The
data are placed in order of decreasing physical scale. 
This ordering is ideal in the sense that a given mode number is associated with
a physical scale, and the signature of a high baryon density (the wiggles) 
are scale dependent.

Computational limits do not allow us to consider the full covariance
matrix for more than $\sim 1000$ Fourier modes, corresponding to $k =
9\times 2\,\pi/L$.  We show, however, that on these scales, the
covariance matrix becomes accurately diagonal, and we use the diagonal
approximation for $n > 728$ modes.  We take care to account for the
correct number of independent modes, using singular value
decomposition for $n < 728$, and a calculation of the effective volume
of our sample for larger $n$.  Calculations are continued to physical
scales of $20h^{-1}$ Mpc, where it is expected that the evolution of
the power spectrum becomes significantly nonlinear.

We have applied this method to several different models, including a
flat $(\Omega_{0}=1)$, and an open $(\Omega_{0}=0.3)$ model, each with
$\Omega_{b}=0.024\pm 0.002h^{-2}$, and a flat model with no baryons.
In the models with baryons, we found that the uncertainties were
inconsistent with $\Omega_{b}=0$ at the $0.8 \sigma$ level in a flat
cosmology, and at the $3.5\sigma$ level in an open cosmology.  The
quantities $\Omega_{b}/\Omega_{0}$, which describes the amplitude of
the wiggles, and $\Gamma$, which describes the overall shape of the
power spectrum, should be close to orthogonal in the fits.  We found
that $\Omega_{b}/\Omega_{0}$ is inconsistent with $\Omega_{b}=0$ at
the $1\sigma$ and $3\sigma$ in the flat and open models, respectively.
Finally, we compared this result with the estimate of uncertainties
which can be derived from band estimates of the power spectrum and
found qualitatively consistent results, all of which lends great
confidence to the idea that in an open universe, $\Omega_{b}$ may be
independently measured from galaxy redshift surveys.

We have ignored in this paper a number of physical effects that need
to be understood before we can claim a definitive calculation of the
power spectrum and its errors.  We plan to use numerical simulations
to include the effects of nonlinearities; if these can be properly
modeled, it is possible that errors on parameters can be tightened
considerably.  Distortions to the density field due to galaxy peculiar
velocities (redshift-space distortions) are known to be an important
effect; including them would complicate the analysis considerably,
although the basis set described in the Appendix may simplify the
problem.  The effects of space curvature and the evolution of
clustering are probably less important, but are straightforward to
include.  Finally, scale- and luminosity-dependent galaxy biasing are
additional complications which are difficult to model realistically,
but it should be straightforward to give rough estimates for how
important these effects are on the estimation of quantities such as
$\Omega_b$.

\acknowledgements{ 
The authors would like to gratefully acknowledge David Spergel,
Michael Vogeley, Max Tegmark and Michael Blanton for invaluable discussions.
We thank David Weinberg for the use of his mock
catalog.  M.A.S. acknowledges support from the Alfred P.
Sloan Foundation. D.M.G. is supported by an NSF Graduate Research Fellowship.
}

\appendix
\section{An Alternate Basis Set}
Throughout the paper, we have used the Fourier modes of the density field as 
our basis set.  However, we could imagine other basis sets that take
the SDSS geometry better into account. 
D. Spergel (private communication)
suggests using 
a form of spherical harmonics and spherical Bessel 
functions for the basis set, with basis vectors given by:
\begin{equation}
\hat{e}_{lmn}=
{\cal Y}_{lm}(\theta,\phi)j_{l}(k_{n}r)\ ,
\end{equation}
where
\begin{equation}
{\cal Y}_{lm}(\theta,\phi)=\left[\frac{2l+1}{4\pi}\frac{(l-m)!}{(l+m)!}
\right]^{1/2}P_{l}^{m}\left(\frac{\cos \theta}{\cos \theta_{0}}
\right)
e^{im\phi}\sqrt{\cos \theta_{0}} \ .
\end{equation}
Here, 
$P_{l}^{m}$ are the Legendre polynomials, ${\cal Y}_{lm}$ are a set of 
functions related to the spherical harmonics,
$j_{l}(x)$ are the spherical Bessel functions, $k_{n}$ is the $n^{th}$ value of
$k$ satisfying $j_{l}(k_{n}R)=0$ at the edge of the survey, 
and $\theta_{0}$ is the radial
angle of the circular cone which defines our space.

It is straightforward to show that this forms a complete, orthonormal
basis set within a circular cone of opening half-angle $\theta_{0}$,
embedded in a sphere of radius $R$.  
The SDSS geometry
is fairly similar.  Therefore, we might imagine that using such a basis
set will give us a very nearly diagonal covariance matrix.  
The resolution scale of these modes are related to $n_{max}$ and
$l_{max}$, the maximum values of $n$ and $l$, respectively 
(see Fisher \etal\ 1995 for a discussion).

We can first decompose the density field into modes of this basis set:
\begin{equation}
\delta({\bf r})=\sum_{l=0}^{l_{max}}\sum_{m=-l}^{l}\sum_{n=1}^{n_{max}}
B_{ln}\hat\delta_{lmn}j_{l}(k_{n}r){\cal Y}_{lm}
(\theta,\phi)\ ,
\end{equation}
where $B_{ln}$ is a 
normalization constant.  The transform conjugate is given by
\begin{equation}
\hat\delta_{lmn}=\int d^{3}{\bf r}\, j_{l}(k_{n}r)
{\cal Y}_{lm}^{\ast}(\theta,\phi)\delta({\bf r})\ .
\end{equation}

Without loss of generality, we can add a separable weighting function 
$\psi({\bf r})=\psi(r)\psi(\theta,\phi)$.
This is included analogously to the weighting function used with the Fourier 
modes, and indeed, the method suggested by Tegmark (1995) is separable in 
radial and angular parts, and thus may provide the best solution to limit
covariance. 

Following Fisher \etal\ (1993), 
the expectation value of the product of any two modes is:
\begin{eqnarray}
\langle \hat{\delta}_{lmn} \hat{\delta}^{\ast}_{l'm'n'}\rangle & = &
\int d^{3}{\bf r}_{1} \int d^{3}{\bf r}_{2}\,\xi({\bf r}_{1}-{\bf r}_{2})
j_{l}(k_{n}r_{1}) j_{l'}(k_{n'}r_{2}){\cal
Y}_{lm}^{\ast}(\theta_{1},\phi_{1}){\cal
Y}_{l'm'}(\theta_{2},\phi_{2})\nonumber \\
& \times & \psi(r_{1})\psi(r_2)\psi(\theta_1,\phi_1) 
\psi(\theta_2,\phi_2)\nonumber \\ 
 & + & \int d^{3}{\bf r} \frac{j_{l}(k_{n}r)j_{l'}(k_{n'}r)
{\cal Y}_{lm}^{\ast}(\theta,\phi){\cal Y}_{l'm'}(\theta,\phi)\psi(r)^{2}
\psi(\theta,\phi)^{2}}{\overline{n}({\bf r})}\ .
\end{eqnarray}
We will label the latter term $C_{lmnl'm'n'}^{shot}$.

Following Fisher \etal\ (1995), we use the Rayleigh expansion to find:
\begin{equation}
e^{i{\bf k}\cdot{\bf r}}=\frac{4\pi}{\cos \theta_{0}}\sum_{l,m}
i^{l}j_{l}(kr\ \cos \theta_{0}){\cal Y}_{lm}^{\ast}(\theta,\phi)
{\cal Y}_{lm}(\theta_k,\phi_k)\ ,
\end{equation}
where $\theta_{k}$ and $\phi_{k}$ are the spherical coordinates of
$\hat {\bf k}$. 

Using this, and the definition of the autocorrelation function, we see:
\begin{eqnarray}
\xi({\bf r}_{1}-{\bf r}_{2})=\frac{2}{\pi\,\cos \theta_{0}}
\sum_{l''m''}{\cal Y}_{l''m''}(\theta_{1},\phi_{1})\psi(\theta_1,\phi_1)
{\cal Y}_{l''m''}(\theta_{2},\phi_{2})\psi(\theta_2,\phi_2)
\nonumber \\
\int k^{2}dk  P(k) j_{l''}(kr_{1}\cos \theta_{0})\psi(r_1)
j_{l''}(kr_{2}\cos \theta_{0})\psi(r_2)\ .
\end{eqnarray}
Plugging this in, we get:
\begin{eqnarray}
C_{lmnl'm'n'}& =& \frac{2}{\pi\cos \theta_{0}}
\sum_{l''m''}A_{lml''m''}A_{l'm'l''m''}^{\ast}
\int dk k^{2}\, P(k)\nonumber \\
& & \int dr_{1} 
r_{1}^{2} j_{l}(k_{n}r_{1})j_{l''}(kr_{1}\cos \theta_{0})
\psi(r_1)\nonumber \\
& & \int dr_{2}
r_{2}^{2} j_{l'}(k_{n'}r_{2})j_{l''}(kr_{2}\cos \theta_{0})\psi(r_2)
+C_{lmn l'm'n'}^{shot}\ ,
\end{eqnarray}
where we have defined: 
\begin{equation}
A_{lml'm'}\equiv \int d\omega\,
{\cal Y}_{lm}^{\ast}(\theta,\phi){\cal Y}_{l'm'}(\theta,\phi)\psi(\theta,\phi)\ ,
\end{equation}

In the case where we have perfect sky coverage over a circle of
radius $\theta_{0}$ and uniform $\psi(\theta,\phi)$, the
$A$ coefficients go to $A_{lml'm'}=\delta^{K}_{ll'}\delta^{K}_{mm'}$, 
which simplifies the form substantially.  
Indeed, since there is only covariance
between different values of $n$, the covariance matrix becomes block diagonal.
Each block can be constructed as a covariance matrix of constant $l$ and
$m$ and varying $n$.  This would simplify storage and inversion tremendously,
allowing us to use all the observed data.  

Perhaps even more importantly, redshift distortions also only couple
different values of $n$ (Fisher \etal\ 1994, 1995; Heavens \& Taylor
1995), and so this formalism should allow these distortions to be
modeled without a great deal of additional complications. 

There are difficulties, however. First, the survey is not perfectly
circular, and hence there is some covariance between different values of
$l$ and $m$.  Moreover, there is a double sum and a triple integral in
this form of the covariance matrix (compared with only a triple integral in
the Fourier picture).  
Moreover, the $A$ coefficients, though not model specific, still need to be
computed for a given geometry, and stored in a four dimensional array.
Finally, computing spherical Bessel functions 
and spherical harmonics are time consuming.

Still, this basis set offers the possibility of an even better approximation
of diagonalizability than the Fourier modes, and may be useful for future
consideration.

\newpage

\begin{figure}
\centerline{\psfig{figure=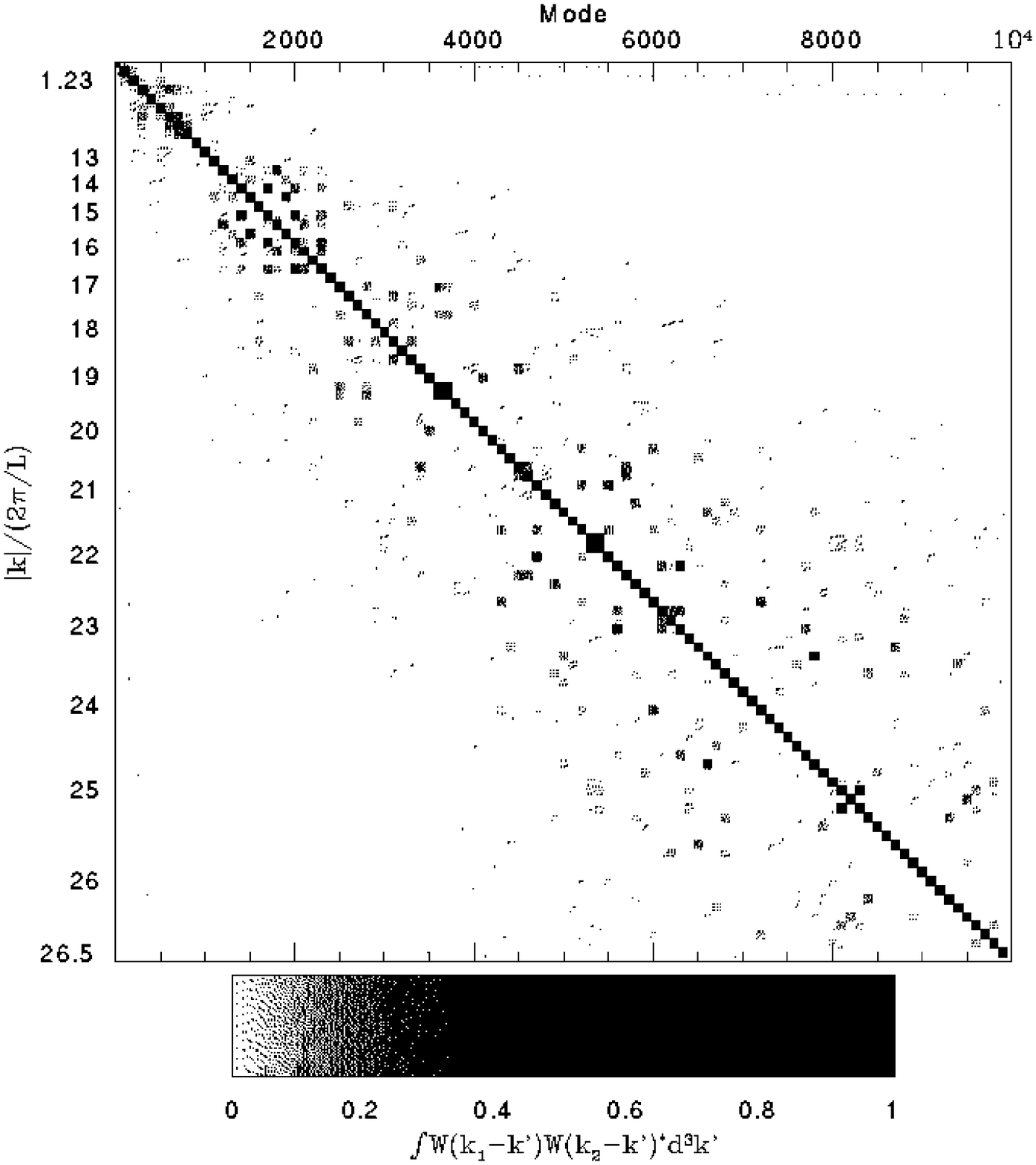,height=6in,angle=0}}
\end{figure}
\begin{figure}
\centerline{\psfig{figure=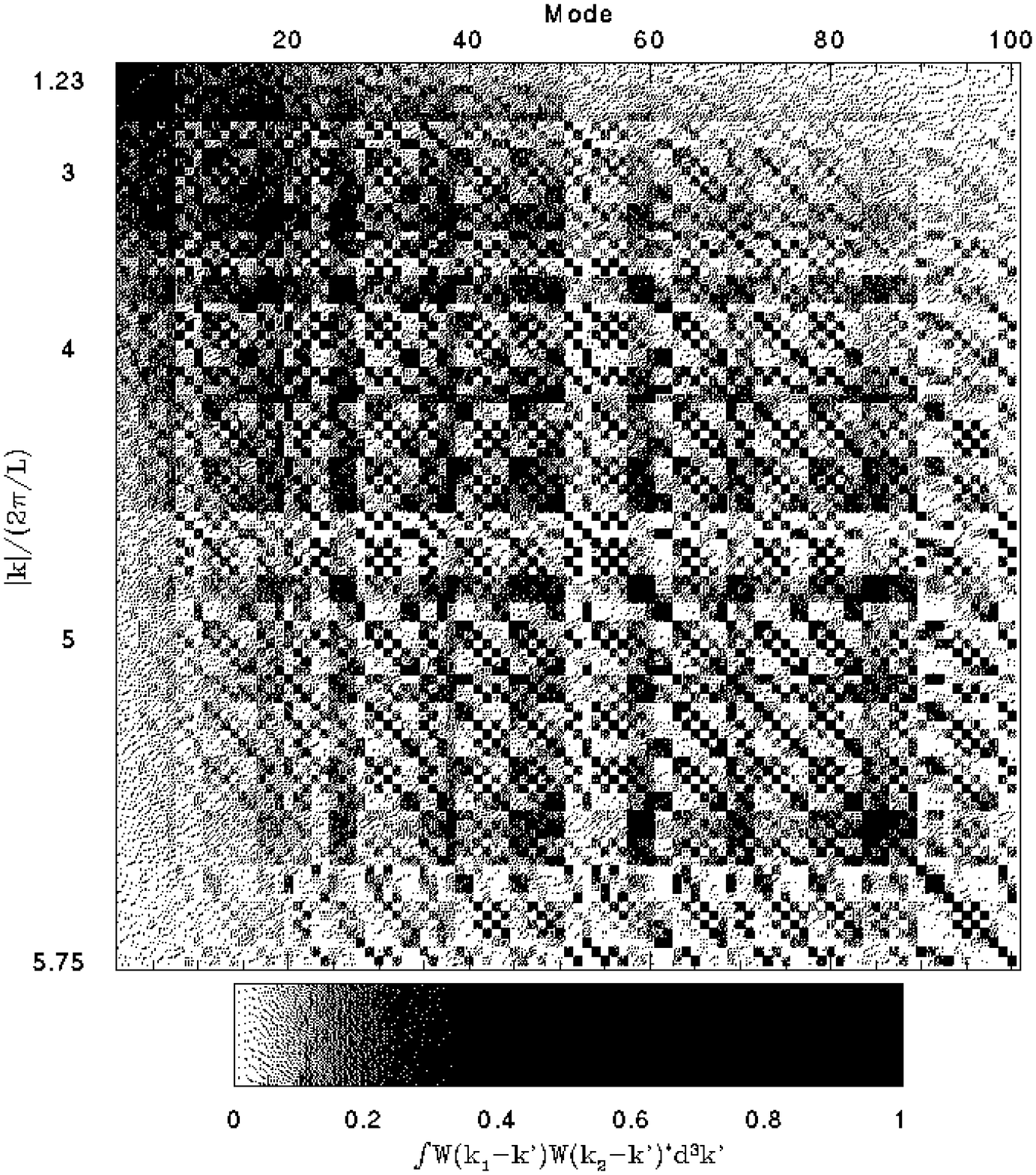,height=6in,angle=0}}
\end{figure}

\end{document}